# Intelligent Traffic Control with Smart Speed Bumps

Melvin Mokhtari*, Amirreza Hosseini†, Alireza Habibi†, Adel Karshenas, Ali Amoomahdi

*Department of Electrical and Computer Engineering, Isfahan University of Technology, Isfahan 84156-83111, Iran*
Email: melvin.mokhtari, amirreza.hosseini, alirezahabibi, adel.karshenas, amoomahdi79@ec.iut.ac.ir

*Abstract*—Traffic congestion and safety continue to pose significant challenges in urban environments. In this paper, we introduce the Smart Speed Bump (SSBump), a novel traffic calming solution that leverages the Internet of Things (IoT) and innovative non-Newtonian fluid materials to enhance road safety, optimize emergency response times, and improve the overall driving experience. The SSBump uses IoT sensors to detect and communicate with emergency vehicles, reducing response times by temporarily deflating. These sensors also analyze traffic patterns and inform data-driven decisions. Additionally, the SSBump uses an Oobleck mixture that adapts its behavior based on the velocity of approaching vehicles, resulting in a safer and more comfortable experience for drivers. This study commences with an overview of the prevalent traffic congestion, followed by a discussion on various available options in this domain. Subsequently, the paper explores the advantages of smart speed bumps and their operational mechanisms. Finally, it presents a comprehensive analysis of the results, its challenges, and the prospects of the work. The findings of this research demonstrate the potential of the SSBump system to revolutionize traffic control, emergency response time, and the driving experience in smart cities, making it a game-changing innovation for advanced transportation systems.

## I. Introduction

In modern urban environments, maintaining safety and efficiency in transportation systems is crucial for the well-being of citizens and the overall functioning of cities. Speed bumps, since their inception in the early 1900s, have been an essential tool for traffic calming and promoting safe driving practices. While conventional speed bumps have remained largely unchanged in their design and materials, emerging technologies and material science advancements offer opportunities to rethink and improve these ubiquitous traffic control devices. This paper introduces the Smart Speed Bump (SSBump), a novel and innovative solution that aims to enhance road safety, optimize emergency response times, and improve the overall driving experience.

It is no secret that conventional speed bumps can be a source of annoyance for drivers, resulting in delays and discomfort. A study has shown that each bump can cause a delay of up to 9.4 seconds [1]. When considering emergency situations, such as a person experiencing a heart attack in an ambulance, a burning building, or a high-speed police chase, every second is invaluable. Consequently, there is a pressing need to minimize delays caused by speed bumps and alleviate any unnecessary inconvenience without compromising road safety.

Moreover, conventional speed bumps have not evolved significantly in terms of design or functionality, despite the wide variety of vehicles that traverse them daily. This lack of customization and innovation presents an opportunity to develop a more versatile and adaptive solution to accommodate different vehicle types and driving scenarios, particularly for emergency services.

The SSBump system leverages the Internet of Things (IoT) and innovative materials to address the limitations of conventional speed bumps. This intelligent traffic control device can operate in various operating modes adapting its behavior according to the specific scenario and vehicle type.

The system is comprised of a network of speed bumps connected via cellular networks to the internet, allowing them to transmit data to our servers. Emergency vehicles are equipped with GPS and LoRa modules, enabling SSBumps to detect their location within the city and communicate with the embedded LoRa modules. This communication is facilitated by LoRa gateways and cellular towers strategically positioned throughout the city, which work in conjunction with software frameworks to process and analyze the collected data.

Data generated by the SSBump system can be converted into wisdom and actionable insights, informing decision-makers and enabling them to develop new policies, road designs, and traffic control strategies. This data can also be used to recommend less congested routes and facilitate traffic flow, which contributes to a faster emergency response time and a more enjoyable driving experience for all road users.

More technically, our system employs GPS sensors to obtain a four-tuple of data, including the unique ID of registered emergency vehicles, their latitude, longitude, and direction of movement. This information is then transmitted to the embedded LoRa modules within the SSBumps. As an emergency vehicle approaches the bump, the system calculates its estimated time of arrival and lowers the height of the bump accordingly. This process occurs within a predefined threshold of seconds, allowing other vehicles to clear the path for the emergency vehicle and reducing fuel consumption and $CO_2$ emissions.

Once the emergency vehicle has passed, the bump seamlessly returns to its original position. This dynamic response ensures that the SSBump remains effective in controlling vehicle speeds while minimizing the impact on emergency response times.

*Correspondence: melvin.mokhtari@ec.iut.ac.ir
†These authors contributed equally.

In some cases, multiple cars pass the system's speed detection sensors before passing the bump. In this case, the system is programmed to consider the speed of the fastest vehicle. This setup has two implications: vehicles moving at a lawful speed experience a smoother ride as the Oobleck material adjusts according to their speed. Meanwhile, fast-moving vehicles not only face a solid bump due to Oobleck's reaction to high-speed objects but also have to navigate a higher-than-usual speed bump.

In addition to the IoT-based architecture shown in Figure 1, the SSBump system also incorporates non-Newtonian fluid materials as an alternative to conventional speed bump materials. Specifically, we chose an Oobleck mixture, a dilatant shear-thickening fluid whose viscosity is proportional to the applied shear rate. This unique property allows the system to adapt its behavior based on the velocity of the approaching vehicle, providing a safer and more comfortable experience for drivers.

When a vehicle encounters the SSBump at a low speed within the critical shear rate of the dilatant fluid, the mixture exhibits liquid-like properties, posing no issues for the vehicle or its occupants. Conversely, if a vehicle impacts the bump at a high speed exceeding the critical shear rate, the fluid behaves as a solid, penalizing reckless drivers.

The combination of the SSBump's adaptable mechanism and the use of non-Newtonian fluid materials makes it the first innovation of its kind, offering a comprehensive solution for future smart cities and intelligent transportation systems.

This paper presents a detailed analysis of the architecture, design, mechanism, and results of SSBump. The analysis shows that SSBump stands out in the market due to the absence of any direct competitors or similar technologies that offer these levels of reliability and efficiency. Specifically, our analysis of the results demonstrates how SSBump has the potential to revolutionize traffic control, emergency response time, and the overall driving experience in smart cities, solidifying its position as a top contender in shaping the future of speed bumps in advanced transportation systems.

## II. RELATED WORK

In previous discussions, it has been noted that no existing system possesses a comparable set of capabilities thus far. However, the current state-of-the-art in the field of intelligent speed bumps remains in its nascent stages, with limited published research. Nonetheless, a few studies have explored the utilization of non-Newtonian fluid materials for speed bump applications, and there have been notable investigations into mechanical lifter bumps.

One study [2] demonstrated that the implementation of viscoelastic fluid asphalts effectively reduced vehicle speeds while simultaneously providing a smoother ride for drivers adhering to the posted speed limit. Another study [3] revealed that magnetorheological fluid brakes offered adjustable resistance levels to vehicle travel, contingent upon the strength of the applied magnetic field. This adaptability allowed for customization of the speed bump to suit specific traffic conditions. Further, a study [4] found that pneumatic lifters provided variable resistance based on the air pressure applied. Additionally, a group [5] undertook a similar project involving a mechanical lifter bump integrated with street-level sensors. They discovered that this approach demonstrated practicality in terms of technical intelligence and reducing emergency response delay, albeit with limitations. Notably, the bump's metallic base material restricted its application to one-lane streets and compromised the smoothness of the ride.

## III. METHODOLOGY

In this section, we present the methodologies and background information that informed our design choices for the Smart Speed Bump (SSBump) system. This discussion covers the various materials, frameworks, and technologies that were considered for the SSBump, as well as the reasons for selecting specific materials and technologies.

### A. Transportation and Networking Layers

There are numerous technologies available for implementing IoT-based solutions in the transportation and networking layers of our system. To select the most suitable technology for our SSBump system, we needed to ensure effective communication between emergency vehicles, speed bumps, and servers. In the following subsections, we analyze the advantages and disadvantages of each technology in detail.

*1) Radio Frequency Identification (RFID):* is a technology that uses readers and tags to transfer data via radio waves, allowing communication distances between a few centimeters and 20 meters. While RFID tags can be read without direct line of sight, they are more expensive, bulkier, and more prone to damage than barcodes. Due to the high communication distance required between the bumps and emergency vehicles, RFID was not chosen for SSBump.

*2) Sigfox:* is a long-range, low-power, low-data-rate wireless communication technology developed for remote sensors, actuators, M2M, and IoT devices. Despite its high communication range, Sigfox's low data rate makes it a less favorable option for SSBump.

*3) Narrowband-IoT (NB-IoT):* is a low-power, wide-area (LPWA) radio technology that connects devices to the IoT using a licensed spectrum [6]. While it provides deep coverage, low-cost hardware, and extended battery life, the use of licensed spectrum increases the costs of constructing and developing SSBump. Consequently, NB-IoT was not chosen for SSBump.

*4) LoRa:* long-range radio is a wireless protocol designed for long-range connectivity and low-power communication in IoT and M2M networks. It enables multi-tenant or public networks to connect various applications and supports millions of nodes per gateway. LoRa also incorporates adaptive data rate algorithms and multiple encryption layers for secure communications. Given its high communication range, sufficient data rate, low cost, and low power consumption, LoRa was selected for use in SSBump to connect emergency vehicles with bumps.

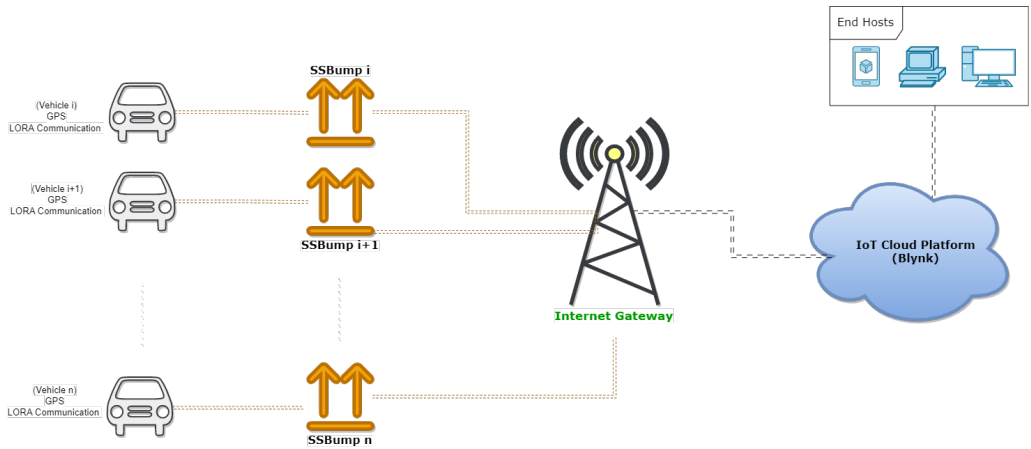

Fig. 1. SSBump Workflow. This is an overview of the SSBump system's workflow and architecture, highlighting the interactions between different entities.

*5) Zigbee:* is an inexpensive, low-power wireless technology designed for machine-to-machine (M2M) networks. It offers low latency, a low duty cycle, and 128-bit AES encryption and is used in mesh networks. However, Zigbee's low communication range makes it unsuitable for SSBump.

*6) Z-Wave:* is an open-source mesh network protocol similar to Zigbee but with slower data throughput and less energy consumption. Like Zigbee, due to Z-Wave's low communication range, it is not suitable for SSBump.

*7) Wi-Fi:* is a medium-range wireless communication system capable of transmitting signals up to 100 meters. However, due to the limited access to Wi-Fi in the city compared to other technologies like 5G, Wi-Fi was not selected for sending data to the servers in the SSBump system.

*8) WiMAX:* is a wireless technology that allows data transfer at a rate of 30–40 megabits per second. While it was once widely used by mobile carriers, it has been largely replaced by faster LTE 4G networks and is no longer a viable option for SSBump.

*9) LTE-M:* is a leading LPWA network technology for IoT applications that connects resource-constrained devices to the internet. It offers low power consumption and high signal penetration, but is not chosen for SSBump due to other available options with better features.

*10) 5G:* is the fifth generation of mobile networks designed to connect virtually everyone and everything, including machines, objects, and devices. It offers higher data speeds, ultra-low latency, increased reliability, network capacity, and availability [7]. 5G is also capable of connecting a massive number of embedded sensors in IoT applications while maintaining low power consumption. Given its numerous advantages, 5G was chosen as the communication technology for connecting bumps with servers in the SSBump system.

*B. Sensors and Actuators*

In order to provide critical decision-making support, we have incorporated features into the SSBump system that enable the use of data in high-level scenarios. To effectively detect the speed and type of approaching vehicles, our system utilizes a well-planned combination of sensors and actuators. In this subsection, we describe the various sensors and actuators considered for the SSBump and explain our choice for each component.

*1) Speed Detection:* To detect the speed of an approaching vehicle, we considered several sensors, including:

- **Radar** sensors emit radio waves that bounce off objects and return to the sensor, allowing the measurement of an object's speed based on the Doppler effect. Radar sensors are widely used in speed measurement applications due to their accuracy and reliability.
- **LiDAR** sensors use lasers to measure the distance to an object and can also calculate its speed based on the time-of-flight principle. While laser sensors offer high accuracy, they are typically more expensive than radars.
- **Ultrasonic** sensors emit sound waves that bounce off objects and return to the sensor, allowing the measurement of an object's speed based on the Doppler effect. Ultrasonic sensors are usually less expensive than radar and LiDAR sensors, but they may be less accurate.

Considering accuracy and reliability, the continuous-wave LiDAR sensor was selected to measure vehicle speeds using the Doppler effect principle, which detects Doppler changes in microwaves reflected from moving objects, and the displacement is proportional to the vehicle speed [8].

*2) Vehicle Type Detection:* To distinguish between vehicles, our system relies on Lora modules, which enable the continuous transmission of emergency vehicle locations to the bumps. Based on this information, distance calculations are performed, and appropriate actions are taken in real-time. While this approach has proven to be highly effective, we recognize the need for a robust backup plan that can minimize errors in our system. As a result, we have explored several alternative detection methods, including:

- **Acoustic sensors** can detect the unique sound signature produced by emergency vehicles' sirens and differentiate them from other vehicles. However, acoustic sensors may be affected by environmental noise and may not be reliable in all conditions.
- **Optical sensors** can analyze the visual appearance of vehicles to identify emergency vehicles based on their markings and light patterns. While this method can be accurate, it may be affected by varying lighting conditions and require complex image processing algorithms.
- **RFID tags** can be attached to emergency vehicles and read by RFID readers installed in the SSBump system. This method provides a reliable and accurate means of identifying emergency vehicles.

Given its high reliability and accuracy, we have opted for RFID technology to serve as the backup system for emergency vehicle detection in the SSBump system. This choice involves the use of RFID tags and readers, which we trust to provide secure and dependable performance in critical situations.

### C. Middleware and Processing Layers

IoT middleware plays a critical role in providing scalability and reliability for systems that handle large numbers of IoT connections and communication loads. Middleware encompasses various components, including data storage modalities, context identification, information extraction, and data reconciliation [9].

Here we will review popular middleware solutions to understand their suitability for the SSBump system.

*1) Microsoft Azure IoT:* is a suite of Microsoft-managed cloud services, edge components, and SDKs, which would be our first choice for the SSBump IoT platform. However, due to the high subscription cost per IoT hub, it was deemed infeasible for our prototype [10].

*2) Amazon AWS:* offers IoT services and solutions designed to connect and manage billions of devices [11]. Similar to Microsoft Azure IoT, Amazon AWS is not suitable for our prototype but could also be considered for the primary SSBump implementation.

*3) Arduino IoT Cloud:* is a platform for creating IoT projects with a user-friendly interface that offers an all-in-one solution for configuration, writing code, and uploading [12].

*4) Blynk:* is another IoT platform for smartphones that enables remote hardware control, sensor data display, and data storage. Blynk's compatibility with various hardware, open-source nature, mobile app and cloud server support, and straightforward user interface make it a better choice and an ideal option for the SSBump project prototype [13].

### D. Non-Newtonian Fluid Materials

As the main surface material for the SSBump, we chose Oobleck, a non-Newtonian fluid created by mixing cornstarch and water in a 2:1 ratio. Oobleck is a dilatant shear-thickening fluid made from a mixture of cornstarch and water [14]. Its unique properties make it an ideal choice for our speed bump system.

Oobleck exhibits the following properties, which make it suitable for use in the SSBump:
- Shear thickening behavior: When exposed to a force, Oobleck's viscosity increases, causing it to act like a solid. This property allows the speed bump to harden when a vehicle passes over it at high speeds, effectively slowing the vehicle down.
- Reversible behavior: Oobleck returns to its liquid state once the applied force is removed. This allows the speed bump to soften when a vehicle passes over it at high speeds, enabling the vehicle to maintain its momentum.
- Environmentally friendly: Oobleck is made from cornstarch and water, making it a non-toxic and biodegradable substance.
- Cost-effective: The primary ingredient of Oobleck, cornstarch, is inexpensive and readily available, which makes it a cost-effective material for the SSBump system.

There are other non-Newtonian fluids that exhibit shear thickening behavior, such as:
- **Polyethylene glycol (PEG) and silica**: This mixture also exhibits shear-thickening properties, but it is more expensive and less environmentally friendly than Oobleck.
- **Shear thickening fluid (STF) armor**: STF armor is composed of liquid-filled packets used in protective clothing and armor applications. While it displays shear-thickening behavior, it is not suitable for the SSBump system due to its complexity and high cost.
- **Dilatant polymer solutions**: Some combinations of polymers and solvents can form dilatant solutions with shear thickening behavior. However, these solutions are usually more expensive and less environmentally friendly than Oobleck.

Given the unique properties of Oobleck and its advantages over other non-Newtonian fluids, we chose Oobleck as the primary material for the surface of the SSBump system.

The utilization of Oobleck, as the surface material for our smart speed bump introduces numerous advantages. In its resting state, the Oobleck surface maintains a semi-solid consistency, offering a smooth driving experience for vehicles adhering to or traveling below the recommended speed limit. However, when confronted by the force of an accelerating vehicle, the Oobleck's non-Newtonian attributes come to the fore, temporarily solidifying to replicate the decelerating effect of a conventional speed bump. This dynamic responsiveness not only showcases innovation but also underscores its environmental sustainability.

Additionally, our smart speed bump incorporates a robust design with a metal base bump upon which the Oobleck layer rests. To contain the Oobleck and ensure longevity, we employ resilient and tear-resistant silicone rubber, known for its flexibility and resistance to extreme temperatures and exposure to the elements.

## IV. PROTOTYPICAL IMPLEMENTATION

In this section, we present the prototype of the proposed SSBump system, with a focus on emergency vehicle response times by dynamically adjusting the height of speed bumps based on the proximity of vehicles. Our prototype consists of three main components: an emergency vehicle, a speed bump replica, and a cloud server.

To implement the prototype, we used several hardware modules. These hardware components and their functionality in the SSBump system are discussed below.

### A. Arduino UNO

This is a popular, beginner-friendly micro-controller based on the ATmega328. Equipped with multiple digital and analog pins, the board can interface with various sensors, actuators, and components. In the SSBump system, the Arduino UNO is employed in both the emergency vehicle and the speed bump. In the emergency vehicle, the Arduino UNO retrieves the current location of the vehicle every 5 seconds and initiates the LoRa module to transmit that information. On the speed bump, the Arduino UNO receives the vehicle speed and the location of the emergency vehicle from the LoRa module and is responsible for adjusting the height of the speed bump accordingly. Figure 2 models this architectural layout.

### B. ESP8266

This is an affordable, well-known WiFi module widely used in Internet of Things projects. As an integrated system-on-a-chip (SoC), it combines a micro-controller unit (MCU) with Wi-Fi capabilities. Due to the size and nature of our prototype, we chose to temporarily use WiFi modules instead of 5G. The ESP8266 functions as an intermediary between the speed bump and cloud servers in the prototype, with the Arduino UNO initiating transmission of data to the servers via WiFi, while 5G technology will be employed in the final product.

### C. LoRa SX1278

This is a long-range, low-power radio capable of communicating over distances up to 10 miles in ideal scenarios. In the SSBump prototype, the SX1278 is used as both a transmitter (on the emergency vehicle) and a receiver (on the speed bump) for sending the location of the emergency vehicle. In our prototype, the module uses a 433 MHz antenna to transmit and receive data.

Figure 3 depicts the architecture of the gateway in the SSBump system.

### D. Other Hardwares

To complete the final product, additional hardware, including a DC power adapter (12V-2.8A), a servo KH42JM2-B01 motor, a microstep driver (9-42V DC), and others, was also utilized. Figure 4 is the final, diminutive prototype of our system.

The prototype codes for the SSBump system are available here.

## V. MARKETABILITY

As urban environments become increasingly connected and intelligent, the demand for innovative traffic control solutions that enhance safety, efficiency, and the overall driving experience is expected to grow exponentially. The SSBump system addresses these needs and presents a promising market opportunity, as no other solutions on the market offer the same levels of reliability, efficiency, and adaptability. Moreover, the increasing emphasis on reducing greenhouse gas emissions and improving air quality will further contribute to the market potential of SSBump, as it reduces fuel consumption and $CO_2$ emissions by minimizing delays and facilitating traffic flow.

The system's ability to adapt to different vehicle types and driving scenarios, in addition to its compatibility with existing infrastructure, makes it an attractive investment opportunity. Furthermore, the data generated by the system can be utilized by various stakeholders, including city planners, transportation departments, and insurance companies, opening up additional revenue streams and business opportunities. All of these factors point to this innovation's promising future.

## VI. RESULTS ANALYSIS

The objective of this section is to examine the notable accomplishments of the SSBump system in contrast to conventional speed bumps, regarding emergency response delay, driver contentment, and overall traffic delay.

*1) Emergency Response Delay:* A critical aspect of the SSBump system is its ability to minimize emergency response delays. Conventional speed bumps have been shown to cause delays of up to 9.4 seconds per bump [1]. The system, leveraging GPS sensors and LoRa communication, eliminates these delays by temporarily deflating the bumps when emergency vehicles approach. This results in zero seconds of delay, which represents a significant improvement over conventional bumps.

*2) Traffic Delays:* We assessed traffic delays caused by regular speed bumps and the SSBump system by comparing road conditions and traffic densities. According to [1], speed bumps are commonly used on roads with speed limits of 40 to 60 km/h, and cars typically slow down to around 12 km/h when passing a regular speed bump. Our observations show that cars begin to decelerate 15 meters after the speed bump and return to normal speed 15 meters after it. This model is shown in Figure 5, which is followed by calculations of the delay for each vehicle when SSBump is used compared to conventional speed bumps.

The analysis reveals the following results:
- Average speed of the car during the 30-meter distance with a regular speed bump: 12 km/h (3.2 m/s)
- Average speed of the car during the 30-meter distance with SSBump: 30 km/h (8.3 m/s)
- Using the formula: time = distance/speed, time to travel 30 meters with a conventional speed bump: 9.4 seconds
- Using the formula: time = distance/speed, time to travel 30 meters with SSBump: 3.6 seconds

According to Table I, the use of the SSBump system results in a significant decrease in the average time delay per vehicle,

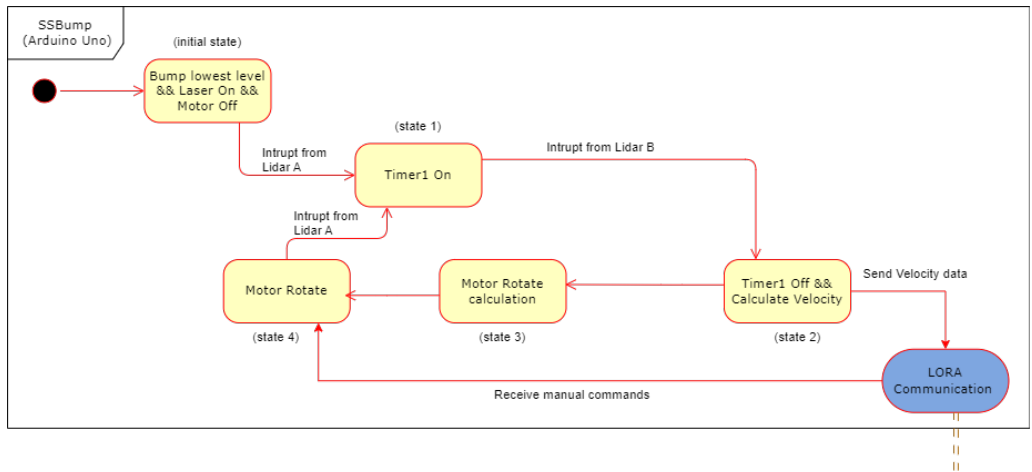

Fig. 2. SSBump State Diagram. This figure highlights the inter-connectivity of bump system elements.

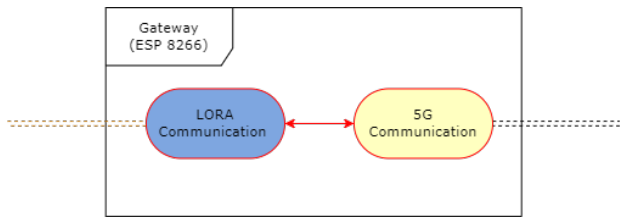

Fig. 3. SSBump Internet Gateway. The SSBump system features a LoRa communication platform on one side for LoRa communications (dotted lines in orange) and a 5G communication platform (WiFi in our prototype) for internet communication (dotted lines in black) on the other side.

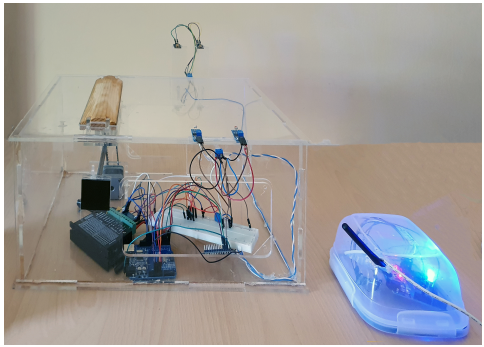

Fig. 4. The SSBump prototype

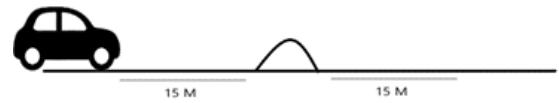

Fig. 5. Traffic delay model

TABLE I
AVERAGE TIME DELAY PER VEHICLE

| Speed Bump Type | Average Delay per Vehicle (seconds) |
| --- | --- |
| Conventional | 9.4 |
| SSBump | 3.6 |

dropping by 61.7%. This reduction in delay has a positive impact on traffic flow and alleviates congestion, especially during peak traffic periods.

*A. Discussion*

The results obtained from the implementation of the Smart Speed Bump system demonstrate its potential to significantly improve traffic control, emergency response times, and the overall driving experience. By incorporating the Internet of Things technology and innovative non-Newtonian fluid materials, the SSBump system addresses the limitations of conventional speed bumps and offers a more versatile and adaptive solution for various vehicle types and driving scenarios.

In the case study presented, the SSBump system was shown to effectively reduce the delay caused by conventional speed bumps, particularly for emergency vehicles. By detecting the approach of vehicles and temporarily deflating the bumps, the system enables faster response times, potentially saving lives and reducing property damage in emergency situations.

Through the use of an Oobleck mixture as a dilatant shear-thickening fluid, the SSBump system provides a more comfortable experience for drivers who adhere to the speed limit. The fluid's unique properties allow the system to adapt its behavior based on the velocity of the approaching vehicle, ensuring a smoother ride for compliant drivers and penalizing those who drive recklessly.

The data generated by the SSBump system can also inform data-driven decisions in various aspects of urban planning and traffic management. For example, by analyzing traffic patterns and congestion levels, city planners can design more efficient road layouts and traffic control strategies. Furthermore, the widespread use and remarkable performance of artificial intelligence and machine learning techniques and models, such as in social incidents, finance [15], and various other fields, enable the prediction of future trends in urban planning. Additionally, the data can be used to recommend less congested routes for drivers, further contributing to a more enjoyable driving experience and a reduction in fuel consumption and CO2 emissions [16].

### B. Challenges

Despite the promising results of the SSBump system, there are some challenges that need to be addressed to ensure its successful implementation and adoption. These challenges include:

*1) Cost and scalability:* The initial investment required to deploy the SSBump system may be relatively high. Balancing the costs with the potential benefits will be crucial for widespread adoption.

*2) Interoperability and standardization:* The integration of the SSBump system with existing traffic control infrastructure and emergency response systems will require the development of common standards and protocols to ensure seamless communication and data exchange.

*3) Privacy and security:* The collection, storage, and transmission of data generated by the SSBump system raises privacy and security concerns. Ensuring the protection of sensitive information and addressing potential vulnerabilities in the system will be critical for user trust and compliance.

### C. Prospects

The SSBump system offers several exciting prospects, both in terms of social and technical advancements:

*1) Social Prospects:*
- **Integration with autonomous vehicle technology**: As autonomous vehicles become more prevalent, the SSBump system could be integrated with these vehicles' navigation systems to provide real-time information on road conditions and traffic patterns and even automatically adapt the speed of vehicles based on their distance to bumps to enhance safety.
- **Integration with smart city initiatives**: The data generated by the SSBump system can be incorporated into broader smart city initiatives, like traffic light systems and public transportation systems, contributing to more efficient resource allocation, better traffic management, and improved quality of life for citizens.
- **Drivers accustomation to the system**: While aligning the speed bump with the speed of cars is a smart approach to traffic control, its long-term effectiveness may diminish as drivers get accustomed to it and start disregarding it. Investigating the psychological and societal factors contributing to this issue and finding potential solutions could be a promising area for future research.

*2) Technical Prospects:*
- **Integration with predictive analytics**: The system can benefit from the use of federated learning and data mining techniques to analyze historical data on traffic patterns, road conditions, and potential hazards, allowing for proactive traffic management and improved safety.
- **Integration with energy harvesting technology**: The development of energy harvesting techniques, such as solar panels or piezoelectric materials, could be used to power the SSBump system, reducing its reliance on external power sources and contributing to its sustainability.

## VII. CONCLUSION

This paper presents the Smart Speed Bump (SSBump) system, a novel and innovative solution to enhance road safety, optimize emergency response times, and improve the overall driving experience in urban environments. By leveraging the Internet of Things and non-Newtonian fluid materials, the SSBump system has demonstrated its potential to revolutionize traffic control and contribute to the development of smart cities and advanced transportation systems.

While the results obtained from the case study are promising, further research is needed to address the challenges associated with cost, scalability, interoperability, privacy, and security. In addition, future work could optimize its performance, enhance its adaptability, and finally pave the way for more significant advancements in urban planning, emergency response, and the overall quality of life in modern cities.